%% file: main.tex
\documentclass[sigconf]{acmart}
\usepackage{xspace}
\usepackage{longtable}
\usepackage{array}
\usepackage[table]{xcolor}
\usepackage{graphicx}

\usepackage{color}
\usepackage{nameref}
\usepackage{multirow}
\usepackage{tabularx} 
\usepackage{censor}
\usepackage{nameref}
\usepackage{enumitem}

\usepackage[most]{tcolorbox}

\usepackage{xcolor}          

\definecolor{lightblue}{RGB}{0, 0, 100}

\newtcolorbox{MyBox}{
  colback=white,
  colframe=lightblue,
  fonttitle=\bfseries,
  coltitle=black,
  sharp corners,
  boxrule=1pt,
  left=5pt,
  right=5pt,
  top=5pt,
  bottom=5pt,
  breakable
}

\definecolor{purplish}{HTML}{D8DFE3}
\definecolor{purplishlight}{HTML}{EBEFF3}
\definecolor{purplishdark}{HTML}{FF7F50}

\newtcolorbox[auto counter,number within=section]{rqbox}[2]{
    nameref=#1,
    title=\small{#1}, 
    enhanced,
    attach boxed title to top left={yshift=-6pt, xshift=8pt},
    boxed title style={size=small,boxsep=1pt},
    colframe=purplishdark,colback=white,colbacktitle=purplishdark,
    boxsep=2pt,left=2pt,right=2pt,top=6pt,bottom=2pt,middle=2pt
}

\AtBeginDocument{%
  }

\setcopyright{acmlicensed}
\copyrightyear{2026}
\acmYear{2026}
\acmDOI{XXXXXXX.XXXXXXX}
\acmConference[WSESE 2026]{}{April 12--18,
  2026}{Woodstock, NY}
\acmISBN{978-1-4503-XXXX-X/2018/06}

\begin{document}

\title{An Investigation on How AI-Generated Responses Affect Software Engineering Surveys}


\author{Ronnie de Souza Santos}
\email{ronnie.desouzasantos@ucalgary.ca}
\orcid{}
\affiliation{%
  \institution{University of Calgary}
  \city{Calgary}
  \state{Alberta}
  \country{Canada}
  }

\author{Italo Santos}
\email{isantos3@hawaii.edu}
\orcid{}
\affiliation{%
  \institution{University of Hawai‘i at Mānoa}
  \city{Honolulu}
  \state{Hawaii}
  \country{USA}
  }

\author{Maria Teresa Baldassarre}
\email{mariateresa.baldassarre@uniba.it}
\orcid{}
\affiliation{%
  \institution{University of Bari}
  \city{Bari}
  \country{Italy}
}

\author{Cleyton Magalhaes}
\email{cleyton.vanut@ufrpe.br}
\orcid{}
\affiliation{%
  \institution{UFRPE}
  \city{Recife}
  \state{Pernambuco}
  \country{Brazil}}

\author{Mairieli Wessel}
\email{mairieli.wessel@ru.nl}
\orcid{}
\affiliation{%
  \institution{Radboud University}
  \city{Nijmegen}
  \country{The Netherlands}
}

\renewcommand{\shortauthors}{XXX et al.}
\begin{abstract}
\input{abstract}
\end{abstract}


\received{20 October 2025}


\maketitle

\newcommand{\searchStrings}{3\xspace}
\newcommand{\topResults}{100\xspace}
\newcommand{\totalResults}{400\xspace}
\newcommand{\initialResults}{100\xspace}
\input{introduction}
\input{background}
\input{method}
\input{findings}
\input{discussion}
\input{conclusion}

\bibliographystyle{ACM-Reference-Format}
\bibliography{bibliography}

\appendix

\end{document}

%% file: abstract.tex
Survey research is a fundamental empirical method in software engineering, enabling the systematic collection of data on professional practices, perceptions, and experiences. However, recent advances in large language models (LLMs) have introduced new risks to survey integrity, as participants can use generative tools to fabricate or manipulate their responses. This study explores how LLMs are being misused in software engineering surveys and investigates the methodological implications of such behavior for data authenticity, validity, and research integrity. We collected data from two survey deployments conducted in 2025 through the Prolific platform and analyzed the content of participants’ answers to identify irregular or falsified responses. A subset of responses suspected of being AI-generated was examined through qualitative pattern inspection, narrative characterization, and automated detection using the Scribbr AI Detector. The analysis revealed recurring structural patterns in 49 survey responses indicating synthetic authorship, including repetitive sequencing, uniform phrasing, and superficial personalization. These false narratives mimicked coherent reasoning while concealing fabricated content, undermining construct, internal, and external validity. Our study identifies data authenticity as an emerging dimension of validity in software engineering surveys. We emphasize that reliable evidence now requires combining automated and interpretive verification procedures, transparent reporting, and community standards to detect and prevent AI-generated responses, thereby protecting the credibility of surveys in software engineering.

%% file: introduction.tex
\section{Introduction}
\label{sec:introduction}
\vspace{-2px}

Large Language Models (LLMs) have become a core element of contemporary software engineering. Their capacity to generate, interpret, and refine source code, documentation, and design artifacts has made them indispensable instruments at multiple stages of the software lifecycle~\cite{fan2023large, kirova2024software, sallou2024breaking, lim2025llm}. Beyond software development support, these systems have begun to shape scientific processes, influencing how research is conducted and reported~\cite{rane2023contribution, liang2024mapping, felizardo2024chatgpt, lecca2025applications}. However, while substantial research has explored the benefits and challenges of incorporating LLMs into software development, discussions about their role in empirical software engineering have progressed more slowly. Recent studies highlight both the promise and the methodological risks of using LLMs to assist data collection, analysis, and reasoning in empirical investigations~\cite{hou2024large, wagner2025towards, lecca2025applications}. Early work has shown that LLMs can improve productivity and support research processes~\cite{rane2023contribution, lecca2025applications}, but has also raised concerns about reproducibility, data leakage, and bias when LLMs are integrated into scientific workflows~\cite{fan2023large, sallou2024breaking, liang2024can, kaiser2025simulating, wagner2025towards}.

Within the empirical software engineering community, particular attention has been paid to the use of LLMs in systematic literature reviews~\cite{huotala2024promise, felizardo2024chatgpt, felizardo2025difficulties, lecca2025applications}. Researchers have applied these models to assist in study selection, screening, data extraction, and qualitative synthesis, with promising results in reducing effort and expanding coverage. Connecting software engineering with other research fields reveals similar discussions about the use of LLMs in various empirical methods, including experiments and studies with human participants~\cite{steinmacher2024can, cui2024can, kaiser2025simulating}. Previous studies show that while LLMs can generate consistent responses and simulate realistic participant behavior, they may also introduce biases or inaccuracies that distort empirical findings~\cite{steinmacher2024can, kaiser2025simulating}. In this expanding body of work, hallucination and misuse emerge as key challenges related to LLM usage in empirical research~\cite{fan2023large, rane2023contribution, liang2024mapping, sallou2024breaking}. Hallucinations compromise the accuracy of generated content, while misuse, intentional or inadvertent, poses serious threats to research integrity. These and other emerging challenges have led to the development of initial guidelines on how to design, document, and report LLM-assisted studies to ensure methodological transparency and reproducibility in software engineering research \cite{wagner2025towards}.

Adding to previous research that examined the challenges and limitations of using LLMs in survey research, experiments, and simulations~\cite{steinmacher2024can, kaiser2025simulating}, \textbf{the present study introduces a new and concerning perspective by analyzing how participants in software engineering surveys intentionally or inadvertently used LLMs to generate, modify, or enhance their responses, resulting in fabricated or manipulated data}. These observations highlight emerging risks to data authenticity and research integrity, demonstrating how generative artificial intelligence (AI) can undermine the validity and trustworthiness of empirical evidence in software engineering. Therefore, this study posed the following overall research question: \textit{How can the use of large language models by participants affect the authenticity and validity of data collected in software engineering surveys?}

To address this question, we conducted a qualitative analysis of AI-generated responses identified in two previously conducted software engineering surveys. We aimed to systematically explore patterns, narrative structures, and stylistic similarities across responses, we characterized how LLM use by participants manifests in survey data, and how such cases can be detected through manual and automated methods. The contribution of this study is threefold: (1) it documents concrete evidence of fabricated and manipulated survey responses within real software engineering surveys, (2) it provides methodological insights into identifying and verifying synthetic narratives through combined qualitative and automated analysis, and (3) it discusses the implications of these findings for data validation, research integrity, and future guidelines in empirical software engineering.

From this introduction, this paper is organized as follows. Section~\ref{sec:background} discusses prior research on the integration of LLMs into empirical software engineering and the methodological concerns raised in recent studies. Section~\ref{sec:method} describes our research design and data collection procedures. Section~\ref{sec:results} presents the observed patterns of LLM misuse in survey participation. Section~\ref{sec:discussion} interprets the implications for research integrity and provides recommendations for prevention. Finally, Section~\ref{sec:conclusion} concludes the paper and outlines directions for future work.

%% file: background.tex
\vspace{-5px}
\section{Survey Research in Software Engineering: Sampling and Validity Concerns} 
\label{sec:background}
\vspace{-2px}

Survey research is a widely used empirical method in software engineering because it enables the systematic collection of data on professional practices, perceptions, and experiences~\cite{kitchenham2002principles, ciolkowski2003practical, molleri2016survey, wagner2020challenges, baltes2022sampling}. A survey extends beyond questionnaires and involves a structured process of defining objectives, designing instruments, recruiting participants, and analyzing responses~\cite{pfleeger2001principles, kitchenham2002principles, molleri2020empirically}. This method is particularly useful for studying phenomena that cannot be directly observed through experiments or case studies, supporting the description of current practices, identifying trends, and evaluating theoretical propositions~\cite{punter2003conducting, ghazi2018survey, wagner2020challenges, molleri2020empirically}. Furthermore, surveys complement experimental and observational studies in building cumulative empirical knowledge \cite{agley2025planning}.

The sampling process plays a key role in survey research because it determines how well the data represent the population of interest~\cite{kitchenham2002principles, de2014sampling, baltes2022sampling}. Foundational guidelines stress that a sampling strategy must align with the survey’s objectives while balancing representativeness, feasibility, and resource constraints~\cite{ciolkowski2003practical, molleri2016survey}. In practice, however, software engineering surveys often rely on convenience, snowball, or self-selected samples obtained through professional networks, academic mailing lists, or conferences~\cite{baltes2022sampling}. These pragmatic solutions increase feasibility, but reduce representativeness and introduce bias, particularly when participants share similar organizational or academic backgrounds~\cite{de2014sampling, ghazi2018survey}.

The absence of well-defined population frames makes sampling particularly challenging in software engineering~\cite{amir2018there}. Reliable information on the number of software professionals worldwide, their distribution across sectors, and their demographic and educational characteristics is rarely available~\cite{wagner2020challenges}. This lack of population data prevents the construction of valid sampling frames and the estimation of sampling error, which restricts statistical generalization~\cite{amir2018there}. Reviews indicate that most studies rely on convenience or self-selected samples and frequently omit essential details about recruitment, non-response, and demographic coverage, which limits the assessment of sampling bias~\cite{molleri2016survey, baltes2022sampling}. Even when surveys achieve large sample sizes, the participants typically represent specific professional or regional communities rather than the field as a whole~\cite{de2014sampling, wagner2020challenges}.

The increasing use of online recruitment platforms has significantly altered survey sampling practices across disciplines~\cite{douglas2023data, agley2025planning}. Platforms such as Prolific~\footnote{\url{https://www.prolific.com}} and Amazon Mechanical Turk~\footnote{\url{https://www.mturk.com}} facilitate large-scale data collection with automated screening, compensation, and management features that make participant recruitment more efficient~\cite{douglas2023data}. These tools are now being used in software engineering to access practitioners beyond conventional academic or professional networks~\cite{alami2024you}. However, verifying participants’ professional identity remains a major concern, as self-reported information does not always correspond to actual training or work experience~\cite{douglas2023data, alami2024you}. Studies highlight the need for layered verification procedures to prevent online samples from including respondents who do not represent the intended population~\cite{douglas2023data, alami2024you}.

These methodological challenges exacerbate long-standing threats to validity in survey research. Construct validity may be weakened by ambiguous wording, inconsistent terminology, or limited alignment between constructs and measures~\cite{pfleeger2001principles, molleri2020empirically}. Internal validity can be influenced by recall bias, social desirability, or inattentive responses, while external validity depends on the representativeness and completeness of the sample~\cite{ciolkowski2003practical, ghazi2018survey, wagner2020challenges}. Online environments introduce additional risks, including automated or duplicate submissions, identity falsification, and participation motivated primarily by financial incentives~\cite{douglas2023data, alami2024you}. Even under strict quality controls, platforms such as Prolific remain vulnerable to misclassified expertise and coordinated response behavior. A recent concern involves the use of LLMs to generate or modify survey responses, creating a new threat to the authenticity and reliability of empirical data~\cite{agley2025planning}. Addressing these risks requires systematic prescreening, piloting, and transparent reporting of recruitment and validation procedures to protect the credibility of survey-based evidence in software engineering~\cite{wagner2020challenges}.

%% file: method.tex
\vspace{-5px}
\section{Methodology}
\label{sec:method}
\vspace{-2px}

During the analysis of two survey datasets, we identified distinct groups of responses that appeared suspicious while performing data validation and qualitative coding. These responses drew attention because of their repetitive structure, similar wording, and uniform composition, suggesting that a subset of participants may have submitted partially or fully generated text rather than authentic answers. This observation served as the initial input that triggered our research question.

To investigate these cases, we followed established guidelines for qualitative studies in software engineering~\cite{seaman1999qualitative, lenberg2024qualitative}. We analyzed the set of responses excluded from the original surveys due to clear evidence of manipulation, transforming what were initially discarded data points into a focused object of inquiry. This approach allowed us to derive empirical insights about the presence and characteristics of AI-generated content in survey datasets, linking methodological reflection directly to the observed patterns reported in our findings.

\vspace{-5px}
\subsection{Survey Contexts}

\textbf{Survey with Software Practitioners.} The first study involved software practitioners recruited through Prolific. Eligibility criteria required professional or recent experience in software development, English fluency, and a minimum age of 18 years. Participants were compensated according to Prolific’s fair payment policy. The survey examined the behavioral and emotional aspects of LLM use in software development, focusing on patterns of reliance, productivity, and self-regulation. It combined closed-ended and open-ended questions across sections on professional background, LLM usage, perceived benefits and risks, and coping strategies. Data were collected over a four-week period and underwent quality control through attention checks, time verification, and manual inspection of open-ended responses. 

\textbf{Survey with Software Engineering Students in the Global South.} The second study explored the experiences, challenges, and career expectations of low-income students enrolled in software engineering programs across the Global South. Recruitment combined convenience, snowball, and the use of Prolific. Platform filters were configured to select students residing in countries of the Global South who were enrolled in higher education and self-identified as low-income. The survey included five open-ended sections addressing motivations, academic challenges, career perspectives, reflections on educational inequality, and demographic information.

\vspace{-5px}
\subsection{Data Collection}
\vspace{-2px}

Both surveys were deployed through the Prolific platform, which enabled participant recruitment, screening, consent collection, and compensation management. Responses were collected anonymously through Qualtrics and exported for validation and analysis. After identifying recurring and unusually similar answers, we isolated all cases showing indicators of manipulation or automation, resulting in a subset of 49 suspicious responses across both surveys. This dedicated dataset was created for the present study. These responses, initially excluded from the primary analyses, became the empirical material for our qualitative reflection.

\vspace{-5px}
\subsection{Analysis}
\vspace{-2px}

The analysis was conducted collaboratively by at least two researchers and consisted of three complementary stages. Following recommendations for qualitative rigor~\cite{seaman1999qualitative, lenberg2024qualitative}, the process combined independent examination, iterative comparison, and triangulation to strengthen validity and reliability. Each phase built upon the previous one, progressing from grounded pattern recognition to interpretive characterization, and finally to cross-validation.

\textbf{Phase 1: Pattern Identification.} Two researchers independently checked all open-ended responses to identify similarities that could indicate artificial generation. Each researcher reviewed the full dataset, marking aspects that demonstrated AI-generated behavior. This process resembled the \textit{constant comparison method}~\cite{seaman1999qualitative}, where categories emerge inductively through repeated confrontation of cases. After an independent review, the researchers compared the observations, discussed the divergences, and agreed on a consolidated set of suspicious characteristics. This comparison promoted inter-coder validation and minimized researcher bias~\cite{lenberg2024qualitative}. 

\textbf{Phase 2: Narrative Characterization.} The same researchers performed a narrative analysis of the suspicious subset to characterize recurrent discursive and emotional structures. Each examined the order of ideas, tone, and linguistic composition to identify rhetorical arcs and stylistic coherence unlikely to appear naturally across participants. This phase employed thematic clustering and cross-case analysis~\cite{seaman1999qualitative} to develop a grounded typology of falsified narratives, with a particular focus on how surface-level coherence obscured the lack of individual nuance. 

\textbf{Phase 3: AI Detection and Triangulation.} Finally, we performed an AI detection validation using the Scribbr AI Detector~\footnote{\url{https://www.scribbr.com/ai-detector/}} to estimate the probability that each suspicious response had been generated or manipulated by an LLM. Automated detection served as a complementary verification technique, a form of \textit{methodological triangulation}~\cite{seaman1999qualitative}, to cross-check qualitative judgments with algorithmic evidence. Among the 49 suspicious responses, 38 (77.6\%) were classified as 100\% AI-generated, seven (14.3\%) as 0\%, and four (8.1\%) as partially AI-generated (14–78\% probability). While these quantitative probabilities supported the manual findings, some ``0\%'' cases showed clear structural and rhetorical similarity to confirmed synthetic responses. Discrepancies were resolved through joint review and interpretive discussion until consensus was reached, consistent with qualitative practices for building dependability through negotiated meaning. This comparison demonstrated how human interpretation and automated detection complement each other, each addressing different levels of evidence and ambiguity.

\vspace{-5px}
\subsection{Ethical and Practical Considerations}
\vspace{-2px}

Both surveys received institutional ethics approval, and informed consent was collected through Prolific. All data were anonymous and stored in secure repositories. In this study, no personally identifiable information was used. The detection of artificial responses raised questions about deception, data validity, and participant compensation. Although excluding falsified submissions preserved the integrity of the original studies, it also revealed the need for clearer participant communication, improved monitoring, and explicit procedures for addressing suspected automation.

\vspace{-5px}
\subsection{Scope, Limitations and Threats to Validity}
\vspace{-2px}

The observations presented here are derived from two survey deployments conducted on the same online platform over a short period. Although similar phenomena may arise in other empirical contexts, these findings are not intended for statistical generalization. Instead, the study offers methodological reflections on how generative models can compromise the authenticity of survey data and outlines practical strategies to detect and mitigate these risks. In documenting these experiences transparently, we aim to inform future survey research in software engineering and support the development of verification procedures that preserve data integrity. Automated verification tools such as the Scribbr AI Detector provide probabilistic estimates rather than definitive classifications, and their performance depends on the text characteristics and model versions they rely on. Consequently, while automated detection supported manual analysis, it cannot fully substitute human interpretive judgment in assessing the authenticity of responses.

%% file: findings.tex
\vspace{-5px}
\section{Findings} \label{sec:results}
\vspace{-2px}

This section presents the main findings derived from the three analytical phases described in our study methodology. Each phase, namely pattern identification, narrative characterization, and AI detection, builds on the previous one to show how falsified or AI-generated responses were detected, clustered, and verified.

\input{tables/clusters}

\vspace{-5px}
\subsection{Detecting: Pattern Identification}
\vspace{-2px}

The first indications of irregularity emerged during the open-coding phase, when both researchers began to experience a recurring sense of \textit{déjà vu} when reviewing open-ended responses. Certain narratives seemed very familiar, as though they had been pre-coded, with repeated expressions and nearly identical openings. To investigate this perception, we revisited the dataset and re-examined the coded material. This process revealed that several responses followed the same progression, often introducing similar contexts, describing comparable experiences, and concluding with analogous reflections.

Once the suspicion was confirmed, we re-examined both survey datasets and verified that a total of 49 responses were falsified or AI-generated. These cases appeared in both studies and consistently showed strong similarities in phrasing, sequencing, and tone. Many reproduced the same rhythm and flow, suggesting that they were derived from similar prompts rather than representing distinct personal accounts. These repetitions occurred even among respondents from different demographic groups or survey conditions, reinforcing that the similarities were not coincidental.

For instance, we observed nearly identical responses from participants in different countries. In Survey 1, a South African participant and another from the United States described their professional use of LLMs using a narrative form and reflective tone that was almost identical. The first wrote that \textit{``while working on a web development project, I encountered a bug in a JavaScript function''} and explained how they \textit{``used ChatGPT to refactor and modularize the function''} through iterative prompting. The second described a parallel situation, stating that \textit{``while working on a data pipeline project, I needed to write a Python script to transform nested JSON data into a flat CSV format''} and that they \textit{``used ChatGPT to generate an initial version of the script.''} Despite referring to different programming tasks, both narratives followed the same pattern: a concise technical problem, a short explanation of how the language model was used, and a reflection on learning or efficiency gains. The identical sequencing and tone across geographically distant participants suggested that both texts were based on the same underlying prompt structure, reinforcing evidence of synthetic authorship within the dataset.

These findings correspond to Phase 1 of our analytical procedure, confirming that pattern identification effectively isolated a subset of synthetic narratives.

\vspace{-5px}
\subsection{Characterizing: Narrative Analysis}
\vspace{-2px}

After fully reanalyzing the 49 responses identified as falsified, we organized them into clusters based on recurring writing patterns observed in the datasets to define the main characteristics of falsified narratives. These clusters demonstrated how the generated responses exhibited consistent forms of organization, tone, and phrasing in both surveys.

\begin{figure}[h]
\centering
\includegraphics[width=1\linewidth]{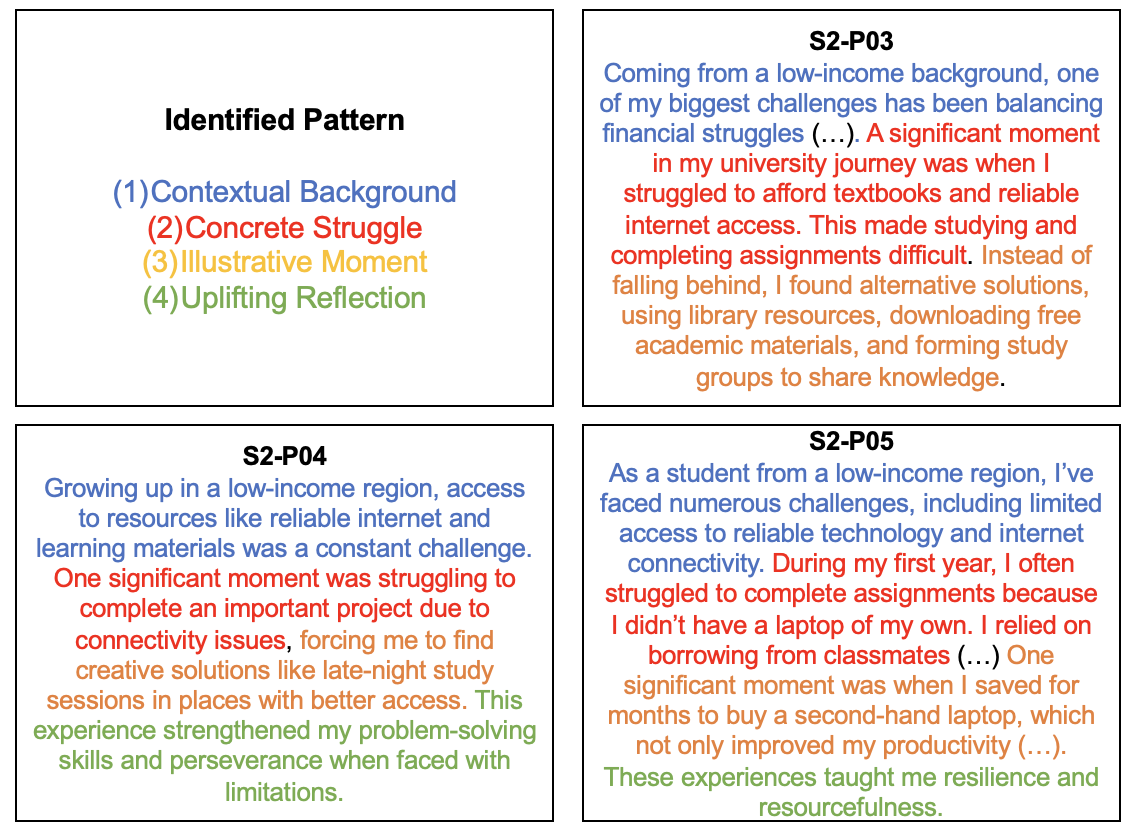}
\caption{Illustrative Pattern Detection}
\label{fig:narrative-pattern}
\end{figure}

A representative example appeared in Survey 2, where three participants (S2-P03, S2-P04, and S2-P05) described their motivations to pursue a career in technology using almost identical arcs. Each began with a contextual statement referencing a \textit{``low-income or underprivileged background''}, followed by a description of a concrete obstacle related to \textit{``limited access to reliable internet, technology, or educational materials.''} They then introduced a supposedly distinct but structurally equivalent \textit{``significant moment''}, such as \textit{``when I struggled to afford textbooks''}, \textit{``when I had to complete a project despite connectivity issues''}, or \textit{``when I saved for months to buy a second-hand laptop.''} All three concluded with a reflection emphasizing resilience and personal growth, using expressions such as \textit{``These experiences taught me resilience and resourcefulness''} or \textit{``They continue to fuel my drive to succeed.''} Across these narratives, the arc followed a four-part sequence: (1) contextual background, (2) concrete struggle, (3) illustrative moment, and (4) uplifting reflection, as shown in Figure~\ref{fig:narrative-pattern}. This structure, repeated with minimal variation, reflected a generated storytelling pattern rather than genuine diversity among participants.

Beyond this example, our analysis revealed additional clusters that displayed similar writing habits in both surveys. In Survey 1, falsified texts were often overly detailed in their task descriptions, reproducing code-related contexts, prompts, and outcomes at a level of uniformity not observed among authentic participants. A similar pattern emerged in Survey 2, where participants used emotionally charged but formulaic storytelling. Their accounts typically opened with a contextual challenge, progressed to a defining event, and concluded with a reflection on resilience or determination.

Although the tone differed between the surveys, technical and procedural in Survey 1, personal and motivational in Survey 2, the organization of the responses remained consistent. In both cases, the grouped texts were coherent and well-written, but showed little variation in expression, depth, or individuality. The clusters of AI-generated responses shared several traits that characterize this type of narrative in surveys:

\begin{itemize}
    \item \textbf{Repetitive Structure:} Narratives followed fixed patterns, with similar sequencing and composition across participants.
    \item \textbf{Uniform Style:} Responses used comparable rhythm and sentence patterns, suggesting generation from shared templates.
    \item \textbf{Reused Phrasing:} Frequent recurrence of expressions such as ``taught me resilience,” “saved me hours,” or “improved efficiency.''
    \item \textbf{Polished but Predictable:} Responses were syntactically correct and logically organized but overly uniform in tone and flow.
    \item \textbf{Superficial Personalization:} While appearing personal, the accounts relied on generic details and lacked the nuance typical of genuine human narratives.
\end{itemize}

In general, these characteristics illustrate how falsified responses reproduced surface-level coherence and emotional cues without the natural variation found in authentic discourse. Table~\ref{tab:narrative-clusters} summarizes the identified clusters.

\vspace{-5px}
\subsection{AI Detection Results}
\vspace{-2px}

After completing the manual comparison, we used the Scribbr AI Detector to estimate the likelihood that an AI model had generated the 49 suspicious responses. The automated results supported our findings. As previously reported, 38 responses (77.6 percent) were classified as fully AI-generated, seven (14.3 percent) as zero percent, and four (8.1 percent) between 14 percent and 78 percent likelihood. Although these results aligned with our qualitative observations, they also revealed limitations of automated detection. Some responses labeled as zero percent AI-generated showed the same structural patterns as those identified as synthetic. A closer inspection suggested that participants had edited or paraphrased the generated text, adding small variations that escaped algorithmic detection but remained evident in manual review. This comparison demonstrates how human and automated analysis complement each other in identifying different levels of manipulation. Table~\ref{tab:survey-narratives} presents representative examples of falsified answers. The complete catalog of the 49 responses, including cross-references and similarity annotations, is available in the supplementary material at \url{https://figshare.com/s/b9b7b6d2d20524881021}
.

\input{tables/results}

\vspace{-5px}
\subsection{RQ. How can the use of large language models by participants affect the authenticity and validity of data collected in software engineering surveys?}
\vspace{-2px}

The use of large language models by participants introduces a new category of data contamination that compromises both the authenticity and validity of the survey findings. Generated or AI-assisted responses can mimic coherent, reflective, and emotionally plausible narratives without representing genuine experiences. This simulation undermines construct validity by distorting the meaning of measured concepts, internal validity by creating artificial consistency across responses, and external validity by blending authentic and synthetic data. Ensuring credible results, therefore, requires the systematic verification of data authenticity through a combination of human and automated detection strategies.

%% file: tables/clusters.tex
\begin{table*}[bt]\scriptsize
\centering
\caption{Clusters of Falsified Narratives Identified Across Both Surveys}
\label{tab:narrative-clusters}
\begin{tabular}{p{2cm}p{2.5cm}p{5.5cm}p{5.5cm}}
\toprule
\textbf{Cluster} & \textbf{Participants} & \textbf{Core Narrative Template} & \textbf{Distinctive Features} \\
\midrule
\textbf{S1 – Cluster 1} & S1-P01, P03, P04, P06, P08, P09, P10, P12, P14, P20, P26, P29, P31, P33, P36, P38, P39 & (1) task framing $\rightarrow$ (2) prompt to LLM $\rightarrow$ (3) model output or code suggestion $\rightarrow$ (4) efficiency or learning reflection & Nearly identical structure describing a programming task, copied prompt, and reflection on time saved or learning gained. \\[0.5em]
\midrule

\textbf{S1 – Cluster 2} & S1-P05, P07, P11, P15, P17, P19, P22, P25, P27, P35, P37 & (1) problem symptom $\rightarrow$ (2) minimal context or snippet $\rightarrow$ (3) LLM explanation or diagnosis $\rightarrow$ (4) fix applied and reflection on productivity & Repeated use of verbs such as \textit{debug}, \textit{identify}, and \textit{improve}, followed by closing reflections on productivity and understanding. \\[0.5em]
\midrule

\textbf{S1 – Cluster 3} & S1-P16, P18, P23, P24, P28, P32, P34, P40 & (1) data or automation task $\rightarrow$ (2) instruction to LLM to generate or refactor script $\rightarrow$ (3) validation or minor edits $\rightarrow$ (4) claim of improved performance & Uniform phrasing describing efficiency improvements, often mentioning “CSV,” “pipeline,” or “data cleaning.” \\[0.5em]
\midrule

\textbf{S1 – Cluster 4} & S1-P02, P21 & (1) query or syntax error $\rightarrow$ (2) LLM troubleshooting prompt $\rightarrow$ (3) diagnostic explanation $\rightarrow$ (4) confirmed fix and reflection & Extremely similar syntax and narrative order, differing only in the specific code or query. \\[0.5em]
\midrule

\textbf{S1 – Cluster 5} & S1-P13, P30 & (1) algorithmic problem $\rightarrow$ (2) LLM refactoring request $\rightarrow$ (3) corrected or optimized version $\rightarrow$ (4) conceptual or efficiency reflection & Shared concise structure focusing on a single coding example and reflection on conceptual clarity. \\[0.5em]
\midrule

\textbf{S2 – Cluster 1} & S2-P03, P04, P05 & (1) contextual background $\rightarrow$ (2) concrete struggle $\rightarrow$ (3) illustrative moment $\rightarrow$ (4) uplifting reflection & Nearly identical motivational narratives describing low-income background, access difficulties, and personal resilience. \\[0.5em]
\midrule

\textbf{S2 – Cluster 2} & S2-P01, P02, P03 & (1) early interest in technology $\rightarrow$ (2) formative person or event $\rightarrow$ (3) decision to pursue program $\rightarrow$ (4) motivational reflection & Similar phrasing and emotional tone, often referencing first exposure to computing and influence of teachers or workshops. \\[0.5em]
\midrule

\textbf{S2 – Cluster 3} & S2-P04 & (1) enumerated criteria $\rightarrow$ (2) repeated justification $\rightarrow$ (3) general concluding statement & Structured as a bulleted-style rationale, presenting selection factors with formulaic phrasing and minimal personalization. \\
\bottomrule
\end{tabular}
\end{table*}

%% file: tables/results.tex
\begin{table*}[t]\scriptsize
\centering
\caption{Survey narratives, matched similarities, and AI detector score}
\label{tab:survey-narratives}
\footnotesize
\begin{tabular}{p{3.3cm} p{1.2cm} p{8cm} p{2cm} p{1.5cm}}
\toprule
\textbf{Survey Question} & \textbf{Participant} & \textbf{Narratives} & \textbf{Similarity} & \textbf{IA Detector} \\
\midrule

Survey 1 - Question: Describe a specific situation when you used an LLM (AI tool) to assist with a coding or programming task at work. &
S1-P16 &
At work, I was tasked with optimizing a Python script that processed large CSV files and frequently ran into memory issues. To solve it, I used an LLM (ChatGPT) to help rewrite the code with a more efficient approach. &
S1-P18, S1-P23, S1-P24, S1-P28, S1-P32, S1-P34, S1-P40 &
100\% \\

 &
S1-P18 &
At work, I was tasked with automating a data-cleaning pipeline for a large set of inconsistent CSV files. I used an LLM (ChatGPT) to generate Python code that could dynamically detect and correct common formatting issues such as inconsistent date formats, null values, and duplicated rows. &
S1-P16, S1-P23, S1-P24, S1-P28, S1-P32, S1-P34, S1-P40 &
100\% \\ \\

 &
S1-P23 &
At work, I was tasked with optimizing a data pipeline written in Python that was processing large CSV files and storing the cleaned data in a PostgreSQL database. The original script was inefficient, taking several minutes to complete due to redundant loops and poor memory management. To assist with this task, I used a Large Language Model (LLM)—specifically, ChatGPT. &
S1-P16, S1-P18, S1-P24, S1-P28, S1-P32, S1-P34, S1-P40 &
78\% \\

\midrule

Survey 2 - Thinking about your journey so far, what aspects of your life experiences led you to choose a Technology program at the University? Was there a specific moment or person who influenced this decision? &
S2-P01 &
Growing up, I was always fascinated by how technology could solve real-world problems. One key moment was when I got access to my first computer and realized how powerful coding could be. I became curious about how software works, which led me to explore programming on my own. A high school teacher who introduced me to coding competitions played a big role in solidifying my interest in technology. &
S2-P02, S2-P03 &
100\% \\ \\
 &
S2-P02 &
``Growing up, I was always fascinated by the endless possibilities that technology offered. From a young age, I spent hours tinkering with gadgets, coding simple programs, and exploring the vast expanse of the digital world.

As I progressed through school, my interest in technology only deepened. I was particularly drawn to the way technology could be used to solve real-world problems and make a tangible impact on people's lives.

One specific moment that solidified my decision to pursue a career in technology was when I attended a coding workshop in high school. The workshop was led by a passionate and knowledgeable instructor who not only taught us the basics of coding but also shared her own experiences of how technology had transformed her life.'' &
S2-P01, S2-P03 &
100\% \\ \\

 &
S2-P03 &
Growing up, I saw how financial struggles impacted my family, and I knew I needed a career that provided stability. I have always been drawn to technology, especially cybersecurity and data analytics, because of its job opportunities and real-world impact. A cybersecurity workshop in high school sparked my passion, and when a family member fell for an online scam, I realized how important this field is. &
S2-P01, S2-P02 &
100\% \\

\midrule

Survey 2 - In difficult moments, what motivates you to continue in the technology field? Is there any kind of support or personal strategy you have developed to overcome these challenges? &
S2-P03 &
``Coming from a low-income background, one of my biggest challenges has been balancing financial struggles with my academic responsibilities. The pressure of knowing my education is not just for me but also a way to uplift my family has been both motivating and stressful.

A significant moment in my university journey was when I struggled to afford textbooks and reliable internet access. This made studying and completing assignments difficult. Instead of falling behind, I found alternative solutions, using library resources, downloading free academic materials, and forming study groups to share knowledge.'' &
S2-P04, S2-P05 &
100\% \\ \\

 &
S2-P04 &
Growing up in a low-income region, access to resources like reliable internet and learning materials was a constant challenge. One significant moment was struggling to complete an important project due to connectivity issues, forcing me to find creative solutions like late-night study sessions in places with better access. &
S2-P03, S2-P05 &
100\% \\ \\
 &
S2-P05 &
``As a student from a low-income region, I’ve faced numerous challenges, including limited access to reliable technology and internet connectivity. During my first year, I often struggled to complete assignments because I didn’t have a laptop of my own. I relied on borrowing from classmates or using shared facilities, which meant late nights at the computer lab to meet deadlines.

These experiences taught me resilience and resourcefulness. I also learned to make the most of free resources like open-source software and online tutorials to fill knowledge gaps. One significant moment was when I saved for months to buy a second-hand laptop, which not only improved my productivity but also reminded me of how perseverance can overcome obstacles. These challenges have shaped me into a more determined, focused student, and they continue to fuel my drive to succeed in my academic and professional journey.'' &
S2-P03, S2-P05 &
100\% \\

\bottomrule
\end{tabular}
\end{table*}

%% file: discussion.tex
\vspace{-5px}
\section{Discussion} 
\label{sec:discussion}
\vspace{-2px}

Our findings illustrate how the methodological vulnerabilities long recognized in survey research \cite{pfleeger2001principles, kitchenham2002principles, ciolkowski2003practical, wagner2020challenges, baltes2022sampling} are being intensified by new forms of data contamination introduced by generative AI. While previous studies have described threats to construct, internal, and external validity related to ambiguous questions, unrepresentative samples, or respondent misrepresentation \cite{ghazi2018survey, molleri2020empirically, douglas2023data, alami2024you}, our results expose an additional and emerging problem: the production of false or AI-generated narratives that imitate human reasoning with linguistic coherence but lack authenticity. 

\textbf{Construct and Internal Validity: Simulated Coherence as False Evidence}. Construct validity in survey research depends on the accurate representation of theoretical concepts through the participants’ genuine experiences \cite{pfleeger2001principles, molleri2020empirically}. The falsified responses identified in both surveys directly challenge this foundation. The repetitive narrative arcs and uniform phrasing in both analyzed surveys demonstrate how generative models can reproduce coherence and emotional tone without corresponding lived experience. These synthetic data create an appearance of validity, where clear narratives hide inaccurate meaning and reduce the clarity of the findings.

Internal validity is also compromised by the artificial consistency across these responses. Authentic participants naturally display variability that reflects personal, contextual, and cognitive differences. The falsified responses, in contrast, exhibit mechanical regularity that can create the illusion of consensus or thematic saturation. Such uniformity risks misleading researchers into inferring shared experiences that do not exist. This situation extends classical internal validity threats such as recall bias or social desirability effects \cite{ciolkowski2003practical, ghazi2018survey} into a new form of algorithmic bias that originates from the generative process itself rather than from human intention. 

\textbf{External Validity: Misrepresentation and Sampling Contamination}. The discovery of falsified responses reinforces ongoing concerns about external validity and representativeness in online data collection. Platforms such as Prolific or Amazon Mechanical Turk allow researchers to access diverse groups of participants, but also introduce risks related to uncertified expertise and inaccurate self-reporting \cite{douglas2023data, alami2024you}. Our findings extend these concerns by showing that misrepresentation now occurs at the level of the response itself. Some participants appear to delegate the generation of their responses to LLMs, transforming authentic respondents into intermediaries for synthetic text. This participation makes it challenging to determine whether the dataset reflects real experiences or a blend of human and generated narratives.

This form of contamination directly affects the interpretability of findings in software engineering, a field that depends on domain-specific expertise for meaningfully contextualized responses. Even when sampling and prescreening procedures follow best practices \cite{baltes2022sampling, de2014sampling}, the authenticity of responses cannot be guaranteed once data are collected. Representativeness, therefore, depends not only on who participates in the survey but also on whether their responses are genuinely self-authored. 

\textbf{Data Authenticity as a Dimension of Validity}. The traditional triad of construct, internal, and external validity remains relevant, yet our findings indicate the necessity of a fourth dimension centered on data authenticity. This concept addresses whether responses genuinely originate from the intended population and reflect authentic cognitive and emotional processes. The falsified narratives identified in our study show that authenticity cannot be assumed, even when research instruments are rigorous and participants are properly recruited. LLMs can emulate human reasoning closely enough to evade conventional quality checks such as timing filters or attention questions~\cite{agley2025planning}. Our comparison of manual and automated detection supports this observation. The detector correctly flagged most falsified texts, but several synthetically authored responses remained undetected because participants had lightly revised the generated text. Verifying authenticity, therefore, requires a combination of human and automated procedures.

\vspace{-5px}
\subsection{Implications for Survey Methodology in Software Engineering}
\vspace{-2px}

Our findings demonstrate that survey research in software engineering now faces both traditional and emerging threats to validity. Established issues, such as poorly defined sampling frames or non-probabilistic recruitment, continue to limit generalizability~\cite{wagner2020challenges}, while AI-mediated participation introduces new risks that affect the content itself. Addressing these challenges requires multi-layered strategies that include prescreening to verify participant eligibility, post-hoc analysis to identify synthetic patterns, and transparent reporting of data-cleaning decisions. Such practices are consistent with calls for methodological rigor and reproducibility in empirical software engineering~\cite{ralph2020empirical}.

Our results also suggest that the definition of valid survey data must evolve. Traditional approaches to data cleaning focus on incomplete or inconsistent responses. In the current context, syntactically perfect but semantically fabricated answers represent a more subtle and potentially more damaging threat. Detecting such content demands interpretive criteria that consider linguistic uniformity, repetitive structure, and lack of contextual depth. As AI assistance becomes increasingly common, methodological standards and reporting guidelines must be updated to protect the integrity of survey-based evidence.

%% file: conclusion.tex
\vspace{-5px}
\section{Conclusions and Future Work}
\label{sec:conclusion}
\vspace{-2px}

This study examined the emerging risks associated with the misuse of LLMs in survey research within software engineering. We investigated cases in which participants employed generative tools to fabricate or manipulate their responses, our study exposed a new category of threat to empirical validity that extends beyond traditional methodological challenges such as sampling, recall bias, or misrepresentation. These findings demonstrate that generative AI can compromise the authenticity of survey data, producing coherent yet fabricated responses that distort evidence and weaken the credibility of empirical conclusions.

Our analysis contributes to a broader understanding of how technological change affects research integrity in software engineering. Although surveys remain an important method for capturing professional practices and perceptions, they now require more attention and adaptation. Ensuring reliable results depends not only on the quality of instrument design or the representativeness othe samples, but also on verifyingion that responses originate from genuine human participants. This perspective introduces data authenticity as a necessary complement to construct, internal, and external validity in contemporary survey research.

Our future work should focus on developing systematic procedures to detect and prevent AI-generated responses in empirical studies. Potential directions include combining automated and interpretive approaches to identify synthetic text, establishing community standards for authenticity verification, and investigating the ethical implications of AI-assisted participation. Advancing these efforts will help protect the integrity of software engineering research, reinforce methodological transparency, and sustain the role of empirical evidence in buildinga reliable body of  knowledge in software engineering.